\documentclass[preprint,aps,superscriptaddress,nofootinbib,tightenlines]{revtex4}
\usepackage{epsfig}
\usepackage{bm}

\def\OMIT#1{{}}

\newcommand{\beq}{\begin{equation}}
\newcommand{\eeq}{\end{equation}}
\newcommand{\bea}{\begin{eqnarray}}
\newcommand{\eea}{\end{eqnarray}}

\newcommand{\benn}{\begin{displaymath}}
\newcommand{\eenn}{\end{displaymath}}

\begin{document}

\begin{figure}[!t]
\vskip -1.5cm
\leftline{
{\epsfxsize=1.8in \epsfbox{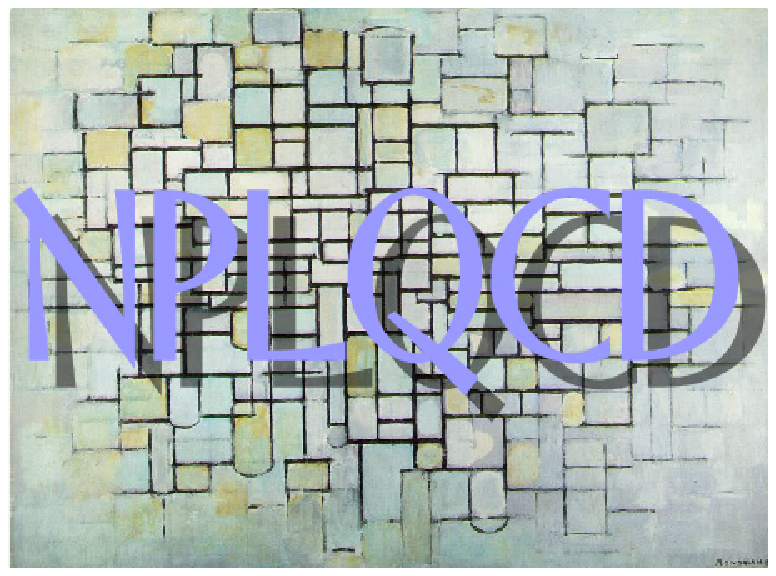}}}
\end{figure}

\preprint{\vbox{
\hbox{NT@UW-06-08}
\hbox{JLAB-THY-06-484}
\hbox{UNH-06-02}
}}

\vphantom{}
\title{\bf \LARGE 
The Gell-Mann -- Okubo Mass Relation among Baryons 
from Fully-Dynamical Mixed-Action Lattice QCD}
\author{Silas R.~Beane}
\affiliation{Department of Physics, University of New Hampshire,
Durham, NH 03824-3568.}
\affiliation{Jefferson Laboratory, 12000 Jefferson Avenue, 
Newport News, VA 23606.}
\author{Kostas Orginos}
\affiliation{Jefferson Laboratory, 12000 Jefferson Avenue, 
Newport News, VA 23606.}
\affiliation{Department of Physics, College of William and Mary, Williamsburg,
  VA 23187-8795.}
\author{Martin J.~Savage}
\affiliation{Department of Physics, University of Washington, 
Seattle, WA 98195-1560.\\
\qquad}
\collaboration{ NPLQCD Collaboration }
\noaffiliation
\vphantom{}
\vskip 0.8cm
\begin{abstract} 
\vskip 0.5cm
\noindent 
We explore the Gell-Mann--Okubo mass relation among the octet baryons
using fully-dynamical, mixed-action (domain-wall on rooted-staggered)
lattice QCD calculations at a lattice spacing of $b\sim 0.125~{\rm
fm}$ and pion masses of $m_\pi\sim$ $290~{\rm MeV}$, $350~{\rm MeV}$, $490~{\rm MeV}$ and $590~{\rm MeV}$.
Deviations from the Gell-Mann--Okubo mass relation are found to
be small at each quark mass.
\end{abstract}
\maketitle

\vfill\eject

\section{Introduction}

\noindent A decade before the formulation of Quantum Chromo-Dynamics
(QCD) as the theory of the strong interactions,
Gell-Mann~\cite{Gell-Mann:1962xb} and Ne'eman~\cite{Ne'eman:1961cd}
proposed that the low-lying hadrons could be embedded into irreducible
representations of the flavor group $SU(3)$, with a fundamental
representation containing three quarks: up, down and strange.  This
provided a relatively simple theory for understanding the plethora of
strongly-interacting particles that had been seen at that time,
although it was a radical proposal as the quarks had not been seen in
isolation.  Now that QCD is known to be the theory of the strong
interaction, classification of the strongly-interacting particles by
their transformation properties under the flavor group $SU(3)$ (and
the isospin subgroup $SU(2)$) follows naturally. Of course, the flavor
symmetries are only approximate, as they are broken by the differences
in the quark masses.  The impact that these mass differences have on
hadronic spectra and interactions is determined by how they compare to
the scale of strong interactions.  The $SU(2)$ symmetry of isospin is
a very-good symmetry of nature, being violated only at the percent
level by both the electromagnetic interactions and the mass difference
between the up and down quarks.  The approximate $SU(3)$ symmetry is
violated typically at the $\sim 30\%$ level due to the large mass
difference between the strange quark and the up and down quarks as
compared to the scale of the strong interaction.  However, there are
instances where $SU(3)$ symmetry works much better than this, as in
the case of matrix elements of an $SU(3)$ charge operator between
states in the same irreducible representation, which are protected by
the Ademollo-Gatto theorem~\cite{Ademollo:1964sr}.

In addition to providing a well-defined scheme for understanding the
spectrum of hadrons, flavor symmetry also provides a vital tool for
the classification of operators, such as those responsible for
nonleptonic decays.  It is well established that the octet component
of the nonleptonic weak interactions dominates over the ${\bf 27}$
component, a feature known as octet dominance.  This is a generic
feature of strong interactions: the contributions from
higher-dimensional representations of $SU(3)$ are suppressed compared
to those from lower-dimensional representations.  The origin of such
suppressions is not well understood, but is found in perturbative
calculations, and also by consideration of the large-$N_C$ limit of
QCD.  The Gell-Mann--Okubo mass relation among the baryons is a
classic example of this feature of QCD.  When all three quark masses
are equal, the light-quark mass matrix transforms as a singlet under
flavor $SU(3)$, and as the strong interactions are also flavor singlets,
higher-dimensional representations of $SU(3)$ are not induced, and
\begin{eqnarray}
M_N & = & M_\Sigma\ =\ M_\Lambda\ =\ M_\Xi
\ \ \ ,
\label{eqn:su3lim}
\end{eqnarray}
where $N=p,n$ denotes the nucleon isodoublet, $\Sigma=\Sigma^+, \Sigma^0,
\Sigma^-$ denotes the $\Sigma$ isotriplet, and $\Xi=\Xi^-, \Xi^0$
denotes the $\Xi$ isodoublet.  In the limit of exact isospin symmetry,
but broken $SU(3)$ symmetry, the light-quark mass matrix takes the
form
\begin{eqnarray}
m_q & = & \left(
\begin{array}{ccc}
\overline{m} & 0 & 0 \\
0 & \overline{m} & 0 \\
0 & 0 & m_s
\end{array}
\right)
\ \ \ ,
\label{eqn:mq}
\end{eqnarray}
where $\overline{m}$ denotes the mass of the up and down quarks, while
$m_s$ denotes the mass of the strange quark. The mass matrix $m_q$
transforms as an ${\bf 8}\oplus {\bf 1}$ under $SU(3)$
transformations.  A single insertion of $m_q$ in an octet baryon mass,
or, more generally, an insertion in the baryon mass that transforms as
an ${\bf 8}$ under $SU(3)$~\footnote{Such a contribution will arise
from $n$ insertions of $m_q$, i.e.  resulting from the ${\bf 8}$'s in
the product $[{\bf 1}\oplus{\bf 8}]\otimes[{\bf 1}\oplus{\bf 8}]\otimes ...\otimes[{\bf 1}\oplus{\bf 8}]$. },
splits the masses of the baryons, but leaves one relation among the
masses intact,
\begin{eqnarray}
T\ \equiv \  M_\Lambda + {\textstyle{1\over 3}} M_\Sigma - {\textstyle{2\over 3}} M_N - {\textstyle{2\over 3}} M_\Xi 
& = & 0
\label{eqn:gmo}
\ \ \ ,
\end{eqnarray}
the Gell-Mann-Okubo (GMO)
mass relation~\cite{Gell-Mann:1962xb,Okubo:1962}.  Multiple insertions
of $m_q$, either through operators in the low-energy effective field
theory, or through loops involving the light mesons, or both, will
lead to violations of this relation.  As $T$ vanishes for $SU(3)$
breaking that transforms as ${\bf 8}\oplus {\bf 1}$, a non-zero
value of $T$ can result only from contributions transforming as 
${\bf 27}$.

The typical scale of $SU(3)$ breaking is set by the differences in the squares
of the meson masses,
\begin{eqnarray}
Q & \sim & { m_K^2-m_\pi^2\over \Lambda_\chi^2}\ \sim\ 0.23
\label{eqn:breaking}
\ \ \ ,
\end{eqnarray}
where $\Lambda_\chi \sim 1~{\rm GeV}$. The mass differences between
baryons in the octet that differ in strangeness by one unit are then
expected to scale as $\Delta M\sim Q \Lambda_\chi$, and one naively
expects violations of the GMO relation to be $T \sim Q^2 \Lambda_\chi
\sim 52~{\rm MeV}$.  Experimentally, at $(m_\pi/f_\pi)^{\rm
expt}=1.045$ , one finds that $T^{\rm expt} = 8.76 \pm 0.08~{\rm
MeV}$~\footnote{The electromagnetic contribution to the violation of
the GMO mass relation is expected to be an order of magnitude less
than $T^{\rm expt}$.}, where the isomultiplet averages of the masses
have been used in eq.~(\ref{eqn:gmo}).  Some theoretical understanding
of why the GMO mass relation works better than naive expectations
would suggest can be gained by considering the large-$N_C$ limit of
QCD, where it is found that violations to this relation are further
suppressed by a factor of $1/N_C$~\cite{Dashen:1994qi} giving $T \sim
Q^2 \Lambda_\chi/N_C \sim 17~{\rm MeV}$, which is close to the
experimental value.
\begin{table}[t]
\label{table:fitting}
\begin{tabular}{|c|c|c|c|c|c|}
\hline
\ $bm_l$ \ & \ $m_\pi$ (MeV) \  & \ $m_\pi/f_\pi$ \  & \  $10^{3}\times T$ (l.u.)\   &\  octet\  centroid (l.u.)\  & \
$10^{3}\times  \delta_{\rm GMO}$\  \\
\cline{1-6}
\cline{1-6}
\ $0.007$\  & $293.1 \pm 1.5$ & $1.983 \pm 0.020$ &  $ 2.4 \pm 1.8 $ & $0.8128 \pm 0.0082$ & $ 3.0\pm 2.3$ \\
\ $0.010$\  & $354.0 \pm 0.9$ & $2.332 \pm 0.012$ &  $ 3.3 \pm 1.0 $ & $0.8218 \pm 0.0080$ & $ 4.1\pm 1.3$ \\
\ $0.020$\  & $492.6 \pm 1.1$ & $3.077\pm 0.014$ &  $ 0.34 \pm 0.56 $ & $0.8724 \pm 0.0055$ & $ 0.39\pm 0.64$ \\
\ $0.030$\  & $591.8 \pm 0.9$ & $3.588\pm 0.011$ &  $ 0.07 \pm 0.42 $ & $0.9162 \pm 0.0041$ & $ 0.07\pm 0.46$ \\
\cline{1-6}  
\end{tabular}
\caption{\it Violations of the Gell-Mann--Okubo mass relation computed with lattice
  QCD. The first and second columns list the sea-quark masses employed in this work
and the corresponding pion masses in MeV, respectively. We have used $b=0.125~{\rm fm}$
to set the scale. The fourth column corresponds to $T$
defined in eq.~(\protect\ref{eqn:gmo}),  measured in lattice units (l.u.),
which is also  the numerator of eq.~(\protect\ref{eqn:gmoratio}).
The fifth column corresponds to the location of the centroid of the
lowest-lying baryon octet (also in lattice units), 
which is the denominator of
 eq.~(\protect\ref{eqn:gmoratio}). All errors quoted are statistical. The systematic
errors due to fitting are all significantly smaller than the statistical errors and thus are not shown.}
\end{table}

Given the smallness of the violation of the GMO relation found
experimentally, it is natural to ask whether this smallness persists
at other values of the light-quark masses, or is no more than an
accident for the physical values of the quark masses.  In this work we
address this issue by computing deviations from the GMO mass relation
at a number of different light-quark masses.  Following the LHP
collaboration (LHPC)~\cite{Edwards:2005kw,Renner:2004ck}, our
computation is a hybrid lattice QCD calculation using domain-wall
valence quarks from a smeared-source on three sets of $N_f=2+1$
asqtad-improved~\cite{Orginos:1999cr,Orginos:1998ue} MILC
configurations generated with rooted-staggered sea
quarks~\cite{Bernard:2001av}. In the generation of the MILC
configurations, the strange-quark mass was fixed near its physical
value, $b m_s = 0.050$, (where $b$ is the lattice spacing) determined
by the mass of hadrons containing strange quarks.  The two light
quarks in the four sets of configurations are degenerate
(isospin-symmetric), with masses $b m_l=0.007, 0.010, 0.020$ and
$0.030$.  Some of the domain-wall valence propagators were previously
generated by LHPC on each of these sets of lattices.  The domain-wall
height is $M_5=1.7$ and the extent of the extra dimension is $L_5=16$.
The parameters used to generate the light-quark propagators have been
``tuned'' so that the mass of the pion computed with the domain-wall
propagators is equal (to few-percent precision) to that of the
lightest staggered pion computed with the same parameters as the gauge
configurations~\cite{Bernard:2001av}. The MILC lattices were
HYP-blocked~\cite{Hasenfratz:2001hp} and Dirichlet boundary conditions
were used to reduce the time extent of the MILC lattices from 64 to 32
time-slices in order to save time in propagator generation.  Various
parts of the lattice were employed to generate multiple sets of
propagators on each lattice.  We analyzed three sets of correlation
functions on 564 lattices with $b m_l=0.030$, 486 lattices with $b
m_l=0.020$, 658 lattices with $b m_l=0.010$, and 468 lattices with $b
m_l=0.007$.  The lattice calculations were performed with the {\it
Chroma} software suite~\cite{Edwards:2004sx,sse2} on the
high-performance computing systems at the Jefferson Laboratory (JLab).
\begin{figure}[!ht]
\vskip 0.46in
\centerline{{\epsfxsize=5.3in \epsfbox{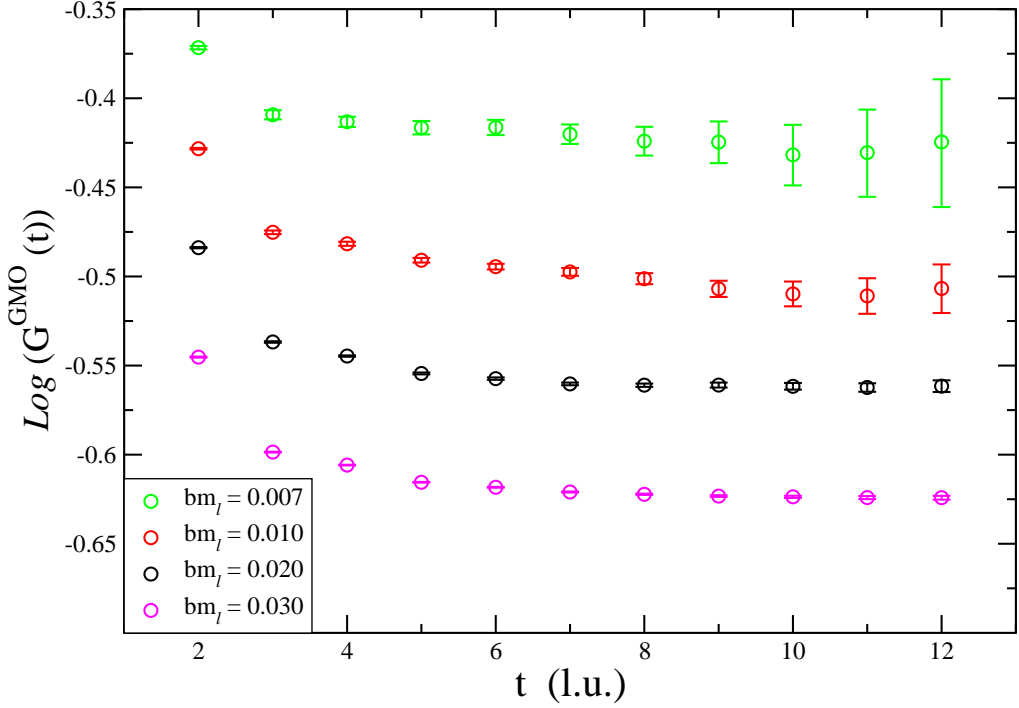}}} 
\noindent
\caption{\it The logarithm of the correlation function $G^{\rm GMO}(t)$ for the
four quark masses employed in this work. Each has been off-set vertically
for display purposes.}
\label{fig:GMOcorrs}
\end{figure}

\begin{figure}[!ht]
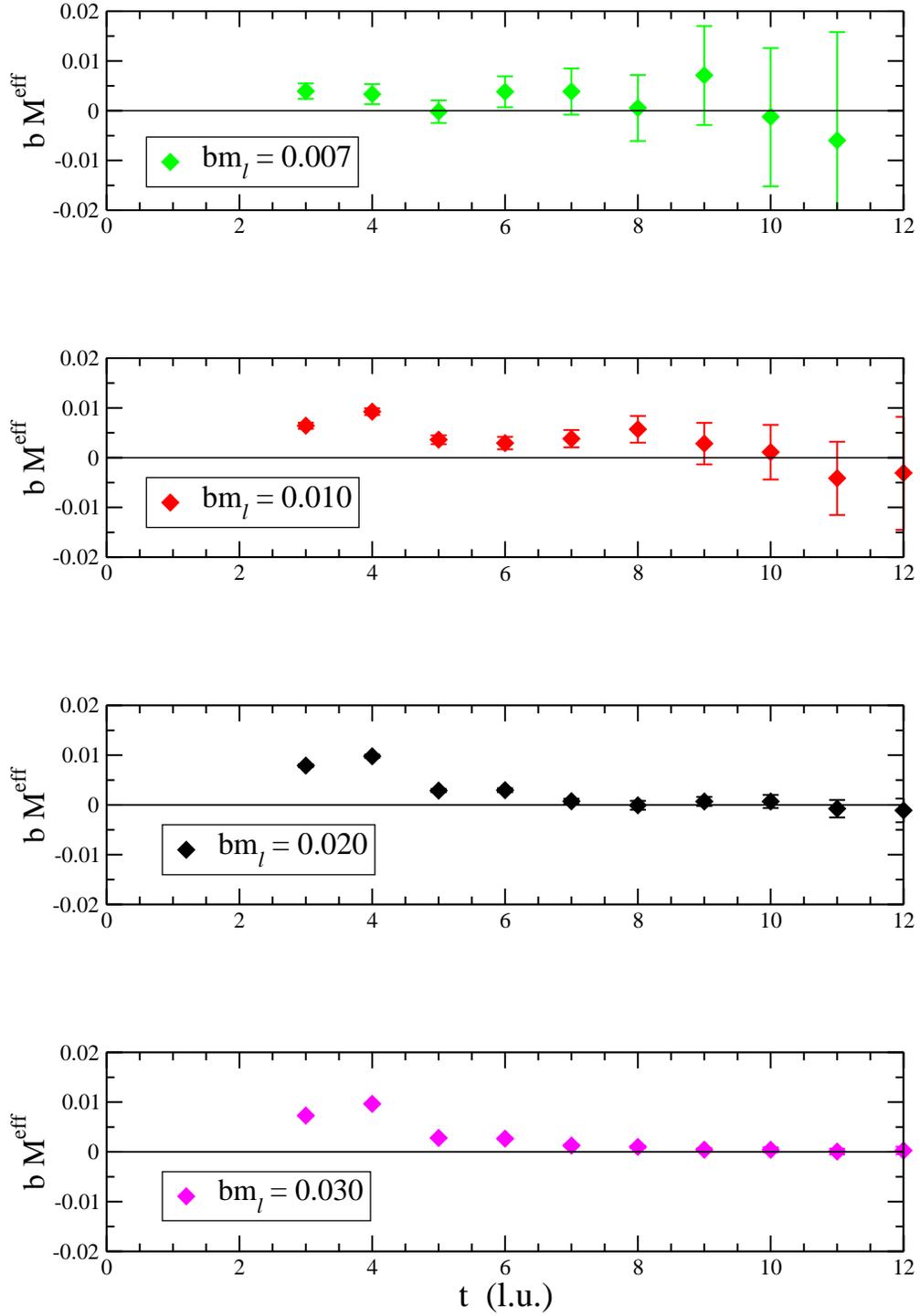

\vskip 0.41in
\centerline{{\epsfxsize=5.2in \epsfbox{EffmassPlot007.eps}}}
\vskip 0.41in
\centerline{{\epsfxsize=5.2in \epsfbox{EffmassPlot010.eps}}}
\vskip 0.41in
\centerline{{\epsfxsize=5.2in \epsfbox{EffmassPlot020.eps}}}
\vskip 0.41in
\centerline{{\epsfxsize=5.2in \epsfbox{EffmassPlot030.eps}}}
\noindent
\caption{\it Effective mass plots for the 
correlation function $G^{\rm GMO}(t)$ at
the four quark masses employed in this work.}
\label{fig:GMOeffmass}
\end{figure}

\begin{figure}[!ht]
\vskip 0.45in
\centerline{{\epsfxsize=5.5in \epsfbox{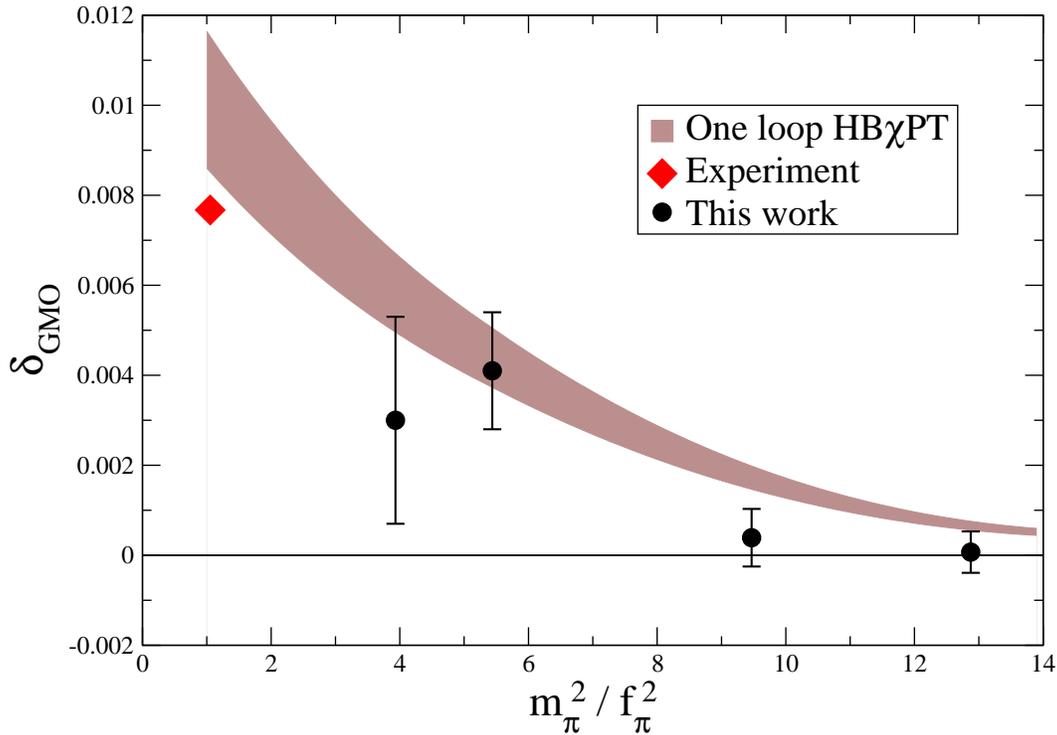}}} 
\vskip 0.15in
\noindent
\caption{\it Deviations from the Gell-Mann--Okubo mass relation for the lowest-lying baryon octet.
The red diamond corresponds to the experimental value. 
The brown band corresponds to the leading-order one-loop 
$SU(3)$ HB$\chi$PT prediction,  and its expected $\sim 30\%$ uncertainty.
The black data points are the results of our fully-dynamical lattice QCD calculation
at pion masses of $m_\pi\sim$ $293~{\rm MeV}$, $354~{\rm MeV}$, $493~{\rm MeV}$ and $592~{\rm MeV}$.
}
\label{fig:results}
\vskip .2in
\end{figure}

We denote as $C_X (t)$ the correlator as a function of time slice for baryon $X$,
with large-time behavior
\begin{eqnarray}
C_X (t) \ \ \rightarrow\ \ e^{- M_X t}\ \ .
\label{eq:bardefined}
\end{eqnarray}
We found that the deviations from the GMO mass relation were
determined most precisely by forming the products and ratios of
correlation functions associated with each of the baryons in the
octet,
\begin{eqnarray}
G^{\rm GMO}(t) & = & { C_\Lambda (t) \ \ \ C_\Sigma (t)^{1/3} 
\over
C_N (t)^{2/3}\ \  C_\Xi (t)^{2/3} }
\ \ \rightarrow\ \ e^{-( M_\Lambda + M_\Sigma /3 - 2 M_N/3 - 2 M_\Xi/3) t } 
\ \ .
\label{eq:ratio}
\end{eqnarray}
The resulting correlation functions are shown in fig.~\ref{fig:GMOcorrs}
and corresponding effective mass plots are shown in fig.~\ref{fig:GMOeffmass}.
It is convenient to define a normalization factor with which to
measure deviations from the GMO mass relation.  We normalize the
deviation to the centroid of the baryon octet, and define the
fractional violation of the GMO mass relation, $\delta_{\rm GMO}$, to
be
\begin{eqnarray}
\delta_{\rm GMO} & = & {M_\Lambda + {\textstyle{1\over 3}} M_\Sigma - {\textstyle{2\over 3}} M_N -
  {\textstyle{2\over 3}} M_\Xi 
\over
{\textstyle{1\over 8}} M_\Lambda + {\textstyle{3\over 8}} M_\Sigma + {\textstyle{1 \over 4}}  M_N + {\textstyle{1\over 4}} M_\Xi }
\ \ \ .
\label{eqn:gmoratio}
\end{eqnarray}
The experimental deviation translates into:
$\delta_{\rm GMO}^{\rm expt}  =  0.00761 \pm  0.00007$.
The results of our fully-dynamical lattice QCD calculation are shown in
Table~\ref{table:fitting}, and are plotted in fig.~\ref{fig:results}.
The computation shows that the deviations from the GMO mass relation
are small, and in fact less than the experimental value, at all quark
masses accessible to us.  It is now quite clear that the small
violation of the GMO relation observed experimentally is not an
accident due to the physical values of the light-quark masses, but is
a generic feature of the baryon spectrum.

Heavy-baryon chiral perturbation
theory~\cite{Jenkins:1990jv,Jenkins:1991ne} (HB$\chi$PT) can be used
to compute corrections to the GMO mass relation.  The leading-order
correction arises at one-loop in the chiral expansion, and is dictated
by the mass differences between the pseudo-Goldstone bosons, the
$\pi$, $K$ and $\eta$.  It is found that~\cite{Bijnens:1985kj,Jenkins:1991ts,Bernard:1993nj}
\begin{eqnarray}
M_\Lambda + {\textstyle{1\over 3}} M_\Sigma - {\textstyle{2\over 3}} M_N -
  {\textstyle{2\over 3}} M_\Xi & = & 
{1\over 24\pi f_\pi^2}\ 
\left[\ 
\left(\ {2\over 3}{\cal D}^2 - 2 {\cal F}^2  \right) \left(\ 4 m_K^3 - 3 m_\eta^3 - m_\pi^3 \ \right)
\right.
\nonumber\\
& & 
\left.\qquad\qquad
- {{\cal C}^2\over 9\pi }\ 
\left(\ 4 F_K - 3 F_\eta - F_\pi\ \right)
\right]
\ \ \ ,
\label{eqn:chiral}
\end{eqnarray}
where the function $F_{c}=F( m_{c},\Delta,\mu)$ is
\begin{eqnarray}
F (m,\Delta,\mu) & = & 
\left(m^2-\Delta^2\right)\left(
\sqrt{\Delta^2-m^2} \log\left({\Delta -\sqrt{\Delta^2-m^2+i\epsilon}\over
\Delta +\sqrt{\Delta^2-m^2+i\epsilon}}\right)
-\Delta \log\left({m^2\over\mu^2}\right)\ \right)
\nonumber\\
& - & {1\over 2}\Delta m^2 \log\left({m^2\over\mu^2}\right)
\ \ \ ,
\label{eq:massfun}
\end{eqnarray}
and we hold the Delta-Nucleon mass-splitting, $\Delta$, fixed at its
physical value, $\Delta^{\rm expt} = 293~{\rm MeV}$~\footnote{The
quark-mass dependence of the decuplet parameters is a higher-order
effect.}. To the order we are working, the $\mu$ independence of the
baryon GMO relation, eq.~(\ref{eqn:chiral}), is protected by the meson
GMO relation, and thus no counterterms are required.  For vanishing
decuplet-octet mass splitting, the function $F$ tends to
$F_P\rightarrow \pi m_P^3$.  The constants ${\cal D}$ and ${\cal F}$
are the axial couplings between the octet baryons and the octet of
pseudo-Goldstone bosons, while the constant ${\cal C}$ is the axial
coupling among the decuplet of baryon resonances, the octet of baryons
and the octet of pseudo-Goldstone bosons.  At tree-level these
couplings take the values~\cite{Butler:1992pn} ${\cal F} = 0.5$,
${\cal D}=0.8$ and ${\cal C}=1.5$, while at one-loop level they take
the values~\cite{Butler:1992pn} ${\cal F} = 0.4$, ${\cal D}=0.6$ and
${\cal C}=1.2$.  In fig.~\ref{fig:results} we have shown (the brown
band) the one-loop prediction of HB$\chi$PT, which is seen to agree,
within uncertainties, with both the experimental value, and also the
results of our lattice calculation. In obtaining this curve, we have
divided the HB$\chi$PT expression of eq.~(\ref{eqn:chiral}) by the
experimental value of the octet centroid. As we have not computed the
mass of the $\eta$-meson on the lattice due to the problems associated
with disconnected diagrams, we have used the Gell-Mann--Okubo mass
relation among the square of the meson masses as input for $m_\eta$
into the one-loop chiral expression in eq.~(\ref{eqn:chiral}), and
used our lattice data for the $\pi$ and $K$ parameters. Expressions
for the octet baryon masses at higher orders in the chiral expansion
have been
computed~\cite{Jenkins:1991ts,Walker-Loud:2004hf,Frink:2004ic}.
However, with lattice calculations at only four pion masses, and the
heaviest likely outside the range of validity of the chiral theory,
there is little point in attempting to fit such expressions.

As with any lattice QCD calculation, there are additional systematic
uncertainties that must be estimated or removed.  The finite-volume
effects on the baryon masses are estimated to be negligible.  The MILC
lattices used in this work have spatial extent $L\sim 2.5~{\rm fm}$
and finite-volume chiral perturbation theory for
baryons~\cite{Beane:2004tw} indicates that for the meson masses used,
the finite-volume effects are very small.  The source of uncertainty
in this calculation that we have the least control over is that due to
the finite lattice spacing, $b$, however such corrections are expected
to be small.  The lattice spacing transforms as a singlet under flavor
transformations, and consequently, it does not contribute at leading
order to violations of the GMO relation.  Even corrections of the form
${\cal O}(m_q b^2)$ will not contribute as they transform as an ${\bf
8}\oplus {\bf 1}$.  However, we expect contributions of the form
$\left(\ 4 m_K^3 - 3 m_\eta^3 - m_\pi^3 \ \right)\ b^2$, which is an
element of the ${\bf 27}$.  Further, there will also be
exponentially-suppressed ${\cal O}( b)$ contributions multiplying
objects that transform as a ${\bf 27}$.  The centroids of the baryon
octet will experience ${\cal O}(b^2)$ shifts, as well as the
exponentially suppressed ${\cal O}( b)$ contributions.  A full
analysis of the chiral corrections at finite lattice spacing should be
performed in the mixed-action theory~\cite{Tiburzi:2005is} when
resources to calculate at another lattice spacing materialize. While
there have been other quenched
QCD~\cite{Bhattacharya:1995ze,Bhattacharya:1995fz,Bowler:1999ae,Aoki:2002fd}
and
full-QCD~\cite{Allton:1998gi,Aoki:2002uc,Aubin:2004wf,Namekawa:2004bi}
studies of octet baryon masses, to our knowledge there have been no
other full-QCD studies of the complete baryon octet in the chiral
regime, prior to this work.

\section{Conclusions}
\label{sec:resdisc}

\noindent 
In this paper we have explored violations of the Gell-Mann--Okubo mass
relation among baryons in the lowest-lying octet using fully-dynamical
lattice QCD. With domain-wall valence quarks on rooted staggered
MILC configurations, we have calculated the deviations to the GMO mass
relation at four different pion masses.  We find that the violations
are consistent with the size of the experimental value, and vanish
toward the limit of exact $SU(3)$ symmetry as required.  Therefore, we
conclude that the smallness of the experimentally-observed violation
of the GMO mass relation is not an accident. This is consistent with
the phenomenological observation that higher-dimensional
representations of the flavor group are suppressed compared with
lower-dimensional representations, and also consistent with the
additional suppression predicted in the large-$N_C$ limit of QCD.

\acknowledgments

\noindent We thank David Kaplan for valuable conversations and Robert
Edwards for help with the QDP++/Chroma programming
environment~\cite{Edwards:2004sx} with which the calculations
discussed here were performed. We are also indebted to the MILC and
the LHP collaborations for use of their configurations and
propagators, respectively. The computations for this work were
performed at the High Performance Computing Center at JLab under the
auspices of the U.S. SciDAC initiative. The work of MJS is supported
in part by the U.S.~Dept.~of Energy under Grant No.~DE-FG03-97ER4014.
The work of KO is supported in part by the U.S.~Dept.~of Energy under
Grant No.~DF-FC02-94ER40818. The work of SRB is supported in part by
the National Science Foundation under grant No.~PHY-0400231.  The work
of KO and SRB is supported in part by DOE contract DE-AC05-84ER40150,
under which the Southeastern Universities Research Association (SURA)
operates the Thomas Jefferson National Accelerator Facility.

\end{document}